\documentclass[ twocolumn, nofootinbib, showpacs,
amsmath, amssymb]{revtex4}

\begin{document}

\title{ Black hole remnants in RG modified gravity? }

\author{ Michael~Maziashvili}
\email{mishamazia@hotmail.com} \affiliation{ Andronikashvili
Institute of Physics, 6 Tamarashvili St., Tbilisi 0177, Georgia
}\affiliation{ Faculty of Physics and Mathematics, Chavchavadze
State University, 32 Chavchavadze Ave., Tbilisi 0179, Georgia }

\begin{abstract}

The idea of black hole remnants for RG modified Schwarzschild solution constructed in papers [hep-th/0002196; hep-th/0602159] depends essentially on the
positiveness of a parameter that enters the running Newton
constant. The positiveness of this parameter was established by
comparing the large distance expansion of RG modified Schwarzschild solution
with the Donoghue's original result about the one-loop correction
to the Newtonian potential [gr-qc/9310024; gr-qc/9405057]. But
since the appearance of paper [gr-qc/0207118] by Khriplovich and
Kirilin it became widely appreciated that the sign of one-loop
correction in Donoghue's original result is incorrect. This
falsifies the argument for existence of black hole remnants in the framework of modified Schwarzschild solution construction suggested in the above papers. But most importantly the very construction of this modified Schwarzschild solution is challenged by the study of graviton radiative corrections to the Newtonian potential and running Newton constant [hep-th/0211071].

\end{abstract}

\pacs{04.60.-m, 04.60.Bc}



\maketitle

Since the mid-1970's black hole remnants were an intriguing subject (see for instance \cite{Ginzburg}) supported in one or another way by the concept of fundamental length (set usually by the Planck length $l_P \simeq 10^{-33}$\,cm) that emerges in many heuristic combinations of general relativity with quantum theory (see for instance one of the first papers on the subject \cite{Mead}). Later on the interest in possible existence of black hole remnants rekindled along the development of various systematic approaches to quantum gravity. One such possibility has been studied in the framework of nonperturbative RG approach to Einstein gravity \cite{BR}. In this brief letter we would like to take a closer look at this discussion. The system of units $c = \hbar =  1$ is assumed in what follows.
The RG modified Schwarzschild metric
\[ ds^2 = -f(r) dt^2 + f(r)^{-1}dr^2 + r^2
d\Omega^2~,
\] was constructed in papers \cite{BR} by using the running Newton constant
\[ G(r)={G_0 \; r^3\over r^3
+ \alpha\,G_0 \, [r+\beta mG_0 ]} ~,\] in a
straightforward way 

\begin{equation}\label{runcoup} f(r)= 1  - {2G(r)m \over r}  ~,\end{equation} where $G_0$ is the macroscopic value of Newton constant
$1/\sqrt{G_0} = l_P^{-1} =  m_P \simeq 10^{19}$\,GeV, the parameter $\beta
\geq 0$, which does not affect very much the horizon structure is
about $4.5$ \cite{BR}. For estimating of $\alpha$ we recall that
in the low energy regime ($\ll m_P$) general relativity can be
successfully treated as an effective quantum field theory
\cite{Donoghue}. So that it is possible to unambiguously compute
quantum effects due to graviton loops, as long as the momentum of
the particles in the loops is cut off at some scale $\ll m_P$. The
results are independent of the structure of any ultraviolet
completion, and therefore constitute genuine low energy
predictions of any quantum theory of gravity. Following this way
of reasoning it has been possible to compute one-loop quantum
correction to the Newtonian potential. Comparing the Eq.(\ref{runcoup})
in the low energy regime, that is, for large values of $r$ with the one-loop corrected Newtonian potential, $V(r)$,

\[ f(r) \approx 1 -  {2 G_0 m \over r}\left(1 - \alpha \,{G_0 \over r^2}\right) = 1 - 2V(r)~, \]
one determines the value of $\alpha$. In paper \cite{BR} the
original result for one-loop quantum correction to the Newtonian
potential \cite{Donoghue, HL}
\begin{equation} \label{old} V(r) = -{G_0 m \over r}\left[1
- {118 \over 15\pi} \,{G_0 \over r^2} \right] ~,\end{equation} was
used for determining of $\alpha$
\[ \alpha = {118 \over 15\pi}~.\] But later on it has been found that the sign of one-loop correction in
Eq.(\ref{old}) is incorrect \cite{Khriplovich}. The correct result
which was checked in several papers \cite{Khriplovich,
Bjerrum-Bohr} strongly indicates the negative value of $ \alpha$.
Taking the revised result \cite{Bjerrum-Bohr}

\[ \label{oneloopcorr}V(r) = -{G_0 m \over r}\left[1 + {41
\over 10\pi} \,{G_0 \over r^2} \right] ~,\] one finds
\[\alpha = - {41 \over 10\pi}~.\]

In the case of positive $\alpha$ the Eq.(\ref{runcoup}) leads to the idea of black hole
remnants as is explained in \cite{BR}. Namely, in this case
(assuming for simplicity $\beta =0$) one obtains two horizons
(outer and inner)

\[ r_{\pm} = G_0m\left[1 \pm \sqrt{1 - {\alpha \over G_0m^2}}\,\,\right]~,
\]that merge together at the critical value of mass \[ m_{cr} = {\sqrt{\alpha \over
G_0}}\equiv \sqrt{\alpha}\,\, m_P~,\]and there is no horizon for
$m < m_{cr}$ \cite{BR}. Hawking temperature is given by the
surface gravity at the outer horizon, \[ T_{\rm H}= {f'(r_{+})
\over 4\pi}= {1\over 4\pi G_0 m}\;{\sqrt{1 - \alpha / G_0m^2}
\over 1 +\sqrt{1 - \alpha / G_0m^2}} ~,
\] so one finds that the Hawking temperature drops to zero when
$m$ approaches $m_{cr}$ that indicates the halt of radiation
\cite{BR}.

For negative $\alpha$ which is actually the case this effect
disappears. Namely, in this case one can easily study the equation
$f(r) = 0$ for determining the horizon structure. This equation
(with arbitrary value of $\beta \geq 0$) reduces to \[ g(x) \equiv
x^3 -2x^2 + \omega x + \beta \omega = 0 ~,\] where $x = r
/G_0m,~\omega = \alpha /G_0m^2$. For $\beta >0$ one easily finds
that as $g(0) = \beta \omega < 0$ and $g(\infty) = \infty$ there
always exists $x_+ > 0$ such that $g(x_+) = 0$. In the case $\beta
 = 0$ too the equation $g(x) = 0$, that is, \[ x^2 -2x + \omega   = 0~, \]
always has a positive root. Thus, one immediately infers that
$g(x)$ function always has at least one positive zero. Thereby the
horizon never disappears that falsifies the argument of \cite{BR}
for existence of black hole remnants.

\vspace{0.4cm}

It should be emphasized that the very construction of modified Schwarzschild solution Eq.(\ref{runcoup}) is challenged by the fact that the one-loop corrected potential \cite{Bjerrum-Bohr} is not obtained by the direct substitution of one-loop running Newton constant \cite{BDH} into the Newtonian potential. This is perhaps the most improtant point that should be aken into account. Indeed comparing the running Newton constan obtained in the RG approach with one-loop result for running Newton constant \cite{BDH}, one finds positive value for $\alpha$ \cite{CPR} but it does not tell us anything about the black hole remnants until we know the correct Schwarzschild solution. Thus the question of paramount importance that naturally occurs in view of the above discussion is to find the proper way for constructing the Schwarzschild solution in RG gravity.

\vspace{0.4cm}

Useful comments from Zurab Silagadze are acknowledged. Author is also indebted to Roberto Percacci for pointing out the paper \cite{BDH} and for useful comments. The work
was supported in part by the \emph{CRDF/GRDF} grant.


\begin{thebibliography}{10}


\bibitem{Ginzburg}

V.~L.~Ginzburg,
  Pisma Zh.\ Eksp.\ Teor.\ Fiz.\  {\bf 22}, 514 (1975).


\bibitem{Mead}

C.~A.~Mead,
  Phys.\ Rev.\  {\bf 135}, B849 (1964);

 C.~A.~Mead,
  Phys.\ Rev.\  {\bf 143}, 990 (1966).


\bibitem{BR}

 A.~Bonanno and M.~Reuter,
  Phys.\ Rev.\  D {\bf 62}, 043008 (2000)
  [arXiv: hep-th/0002196];


A.~Bonanno and M.~Reuter,
  Phys.\ Rev.\  D {\bf 73}, 083005 (2006)
  [arXiv: hep-th/0602159].




\bibitem{Donoghue}


J.~F.~Donoghue,
  Phys.\ Rev.\ Lett.\  {\bf 72}, 2996 (1994)
  [arXiv: gr-qc/9310024];


J.~F.~Donoghue,
  Phys.\ Rev.\  D {\bf 50}, 3874 (1994)
  [arXiv: gr-qc/9405057].


\bibitem{HL}

 H.~W.~Hamber and S.~Liu,
  Phys.\ Lett.\  B {\bf 357}, 51 (1995)
  [arXiv: hep-th/9505182].






\bibitem{Khriplovich}

 I.~B.~Khriplovich and G.~G.~Kirilin,
  J.\ Exp.\ Theor.\ Phys.\  {\bf 95}, 981 (2002)
  [Zh.\ Eksp.\ Teor.\ Fiz.\  {\bf 95}, 1139 (2002)]
  [arXiv: gr-qc/0207118].


\bibitem{Bjerrum-Bohr}

 N.~E.~J.~Bjerrum-Bohr, J.~F.~Donoghue and B.~R.~Holstein,
  Phys.\ Rev.\  D {\bf 67}, 084033 (2003)
  [Erratum-ibid.\  D {\bf 71}, 069903 (2005)]
  [arXiv: hep-th/0211072];


A.~Akhundov and A.~Shiekh,
  Electron.\ J.\ Theor.\ Phys.\  {\bf 17}, 1 (2008)
  [arXiv: gr-qc/0611091];



 S.~Faller,
  Phys.\ Rev.\  D {\bf 77}, 124039 (2008)
  [arXiv: 0708.1701 [hep-th]].



 I.~B.~Khriplovich and G.~G.~Kirilin,
  J.\ Exp.\ Theor.\ Phys.\  {\bf 98}, 1063 (2004) [Zh.\ Eksp.\ Teor.\ Fiz.\  {\bf 125}, 1219 (2004)]
  [arXiv: gr-qc/0402018].


\bibitem{BDH}

 N.~E.~J.~Bjerrum-Bohr, J.~F.~Donoghue and B.~R.~Holstein,
  Phys.\ Rev.\  D {\bf 68}, 084005 (2003)
  [Erratum-ibid.\  D {\bf 71}, 069904 (2005)]
  [arXiv: hep-th/0211071].


\bibitem{CPR}


A.~Codello, R.~Percacci and C.~Rahmede,
  Annals Phys.\  {\bf 324}, 414 (2009)
  [arXiv: 0805.2909 [hep-th]].


\end{thebibliography}
\end{document}